\documentclass[pra,twocolumn,showpacs,amsmath,amssymb,superscriptaddress]{revtex4-1}
\usepackage{psfrag,graphicx}
\usepackage{amsfonts,amssymb,amsmath}      
\usepackage{graphicx}
\usepackage{dcolumn}
\usepackage{bm}
\usepackage{float}
\usepackage{hyperref}
\usepackage{color,soul}
\usepackage{epstopdf}

\newcommand{\nn}{\nonumber\\}
\newcommand{\ex}[1]{\langle{#1}\rangle}

\newcommand{\gam}{\mu}

\newcommand{\erf}[1]{Eq.~(\ref{#1})}

\newcommand{\beq}{  \begin{equation} }
\newcommand{\eeq}{\end{equation}}

\definecolor{nblue}{rgb}{0.06,0.3,0.73}
\definecolor{nblack}{rgb}{0,0,0}

\newcommand{\blk}{\color{nblack}}

\begin{document}

\title{Adaptive estimation of a time-varying phase with coherent states: smoothing can give an unbounded improvement over filtering}

\author{Kiarn T. Laverick}
\affiliation{Centre for Quantum Computation and Communication Technology 
(Australian Research Council), \\ Centre for Quantum Dynamics, Griffith University, Nathan, QLD 4111, Australia}
\author{Howard M. Wiseman}
\affiliation{Centre for Quantum Computation and Communication Technology 
(Australian Research Council), \\ Centre for Quantum Dynamics, Griffith University, Nathan, QLD 4111, Australia}
\author{Hossein T. Dinani}
\affiliation{Facultad de F\'isica, Pontificia Universidad Cat\'olica de Chile, Santiago 7820436, Chile\\} 
\affiliation{Department of Physics and Astronomy, Macquarie University, Sydney, NSW 2109, Australia\\} 
\author{Dominic W. Berry}
\affiliation{Department of Physics and Astronomy, Macquarie University, Sydney, NSW 2109, Australia\\}

\date{\today}

\begin{abstract}
The problem of measuring a time-varying phase, even when the statistics of the variation 
is known, is considerably harder than that of measuring a constant phase. In particular, 
the usual bounds on accuracy --- such as the $1/(4\bar{n})$ standard quantum limit with coherent states --- 
do not apply. 
Here, restricting to coherent states, we are able to analytically obtain the achievable accuracy --- the equivalent of the standard quantum limit --- for a wide class of phase variation.  
In particular, we consider the case where the phase has Gaussian statistics and a power-law spectrum equal to 
$\kappa^{p-1}/|\omega|^p$ for large $\omega$, for some $p>1$. 
For coherent states with 
mean photon flux ${\cal N}$, we give the 
Quantum Cram{\'e}r-Rao Bound on the mean-square phase error 
as $[p \sin (\pi/p)]^{-1}(4{\cal N}/\kappa)^{-(p-1)/p}$.
Next, we consider whether the bound can be achieved by 
an adaptive homodyne measurement, in the limit 
${\cal N}/\kappa \gg 1$ which allows the photocurrent to be linearized. 
Applying the optimal filtering for the resultant linear Gaussian system, we find the same scaling with ${\cal N}$, but with
a prefactor larger by a factor of $p$. 
By contrast, 
if we employ optimal smoothing we can exactly obtain the Quantum Cram{\'e}r-Rao Bound. 
That is, contrary to previously considered ($p=2$) cases of phase estimation, here the improvement 
offered by smoothing over filtering is not limited to a factor of 2 but rather can be unbounded by a factor of $p$. We also study numerically the performance of these estimators for an adaptive measurement in the limit where ${\cal N}/\kappa$ is not large, and find a more complicated picture.
\end{abstract}

\maketitle

\section{Introduction}
Estimating a phase imposed on an optical beam is an important task in quantum metrology, with applications in many areas~\cite{WisMil10}. 
Here we consider a phase shift on a single beam, which is estimated via `dyne' 
measurements~\cite{WisMil10}. That is, the phase is measured relative to a strong local oscillator (LO), which is treated classically, and only the intensity in the beam carrying the phase information is considered as the resource.
Standard techniques use coherent states, 
and the accuracy is limited due to the shot noise of coherent states. 
The limit for coherent states is called the standard quantum limit (SQL).
Alternatively one may use squeezed states or more advanced states to improve the accuracy, 
as originally proposed in Ref.~\cite{Caves81}.
The ultimate limit to the accuracy using arbitrary states is often called the Heisenberg limit.

Phase measurements are most easily analyzed when the phase is constant.
In that case, the resource is just the average photon number $\bar n$. 
In the limit $\bar n \gg 1$, 
the 
SQL on the mean-square error (MSE) is proportional to $1/\bar n$ \cite{Leon95}, and the Heisenberg limit is proportional to $1/\bar n^2$ \cite{Pegg90}.
There was much debate over the ultimate limits to phase measurement \cite{Bollinger96,Yurke86,Sanders95,Ou96,Zwierz10,Rivas12,Luis13,Luis13b,Anisimov10,Zhang13}, but rigorous proofs now exist~\cite{Tsang12,Gio121,Berry12,Hall12,Nair12,Gio122,Hall12b}. 

In many applications, the phase varies continually in time, so the above 
results do not apply. 
In this situation, the appropriate resource is not the mean photon number 
(which depends on the integration time of the measurement) but 
rather
the average photon flux, ${\cal N}$.
To analyze this problem, it is necessary to consider a particular form of variation for the phase.
Early work considered phase that varies as a Wiener process, and analyzed adaptive measurements on squeezed beams~\cite{DominicPRA02,DominicPRA06,DominicE06}.
Later work considered the more general case of Gaussian phase variation with a spectrum scaling as $1/|\omega|^p$ for $p>1$, and derived ultimate (Heisenberg) limits on the accuracy~\cite{DominicPRL13,DominicPRX}. 
Recently it was shown how to achieve the same scaling as the ultimate limit using adaptive measurements on squeezed beams, albeit with a different prefactor \cite{DinaniARX}.

Reference \cite{DominicPRL13} also considered the coherent state case, and derived a scaling proportional to $1/{\cal N}^{(p-1)/p}$ for a rigorous lower bound on the mean-square phase error. This Quantum Cram{\'e}r-Rao Bound (QCRB) for coherent states can 
be regarded as a SQL for a varying phase. It is the coherent state case 
with which we are concerned in this paper. 

Here, we obtain the prefactor in the QCRB for an optimal unbiased measurement.
We then show, using optimal filtering, that the QCRB scaling can be achieved for a phase estimate, however the prefactor will never be as low as that in the QCRB.
Interestingly, if we consider a spectrum scaling as $1/|\omega|^p$ for the phase, the prefactor compared to the QCRB grows unboundedly by a factor of $p$. However, as we go on to show, the technique of {\em smoothing}~\cite{TSL09,vanTrees1,vanTrees2,Wheatley,Yonezawa,Weinert01}  does allow the lowest possible MSE --- that of the QCRB --- 
to be achieved, for an {\em arbitrary} phase spectrum. 
If we then consider a power law spectral density for the phase we can model the system as a linear Gaussian (LG) estimation problem.
By doing so we can study the convergence, as ${\cal N}$ increases, of the performance of the optimal linear filter and smoother to its asymptotic value, for $p$ an even integer. We do this for $p=2$ and $p=4$, 
and also numerically demonstrate for $p=4$ that a suboptimal filter that has 
previously been employed in many theoretical 
treatments~\cite{DominicPRA02,DominicPRA06,DominicE06} fails to converge. 
For the optimal smoother, we confirm numerically that the advantage over filtering in terms of MSE is a factor of $4$ for $p=4$. This surpasses the 
factor of $2$ previously observed~\cite{TSL09,Wheatley,Yonezawa} for $p=2$ and an 
unbounded improvement is predicted as $p$ increases. 

First, in Sec.~\ref{sec-qcrb} we will discuss the type of system we will consider and review Fisher information to find the QCRB. In Sec.~\ref{sec-optf}, having defined the problem, we apply Wiener filtering to find the error. Next we apply Wiener smoothing to attain the QCRB in Sec.~\ref{sec-smooth}. We then model this system as a LG system in Sec.~\ref{sec-LG}. Finally, we will simulate this system without linearizing the photocurrent and compare it to the linearized results in Sec.~\ref{sec-num}, so as to explore 
the regime of low intensity. 

\section{Quantum Cram\'{e}r-Rao Bound}
\label{sec-qcrb}
Consider a time-varying phase $\varphi(t)$ for a coherent beam with Gaussian and stationary statistics. 
This means that the expectation value of the phase $\ex{\varphi(t)}$ is time-independent and can be taken to be equal to $0$.
Furthermore, the autocorrelation function $\Sigma(t,t')=\ex{\varphi(t)\varphi(t')}$ will only be a function of $t-t'$ and hence can be expressed using a single time argument $t$.
The spectral density $S(\omega)$ is given by the Fourier transform of the autocorrelation function, 
\beq
S(\omega)=\int_{-\infty}^\infty dt \, \Sigma(t) \, e^{-i\omega t} \, .
\eeq

The QCRB for coherent states is derived using Fisher information, 
following the approach in Ref.~\cite{DominicPRL13}, based 
on 
Ref.~\cite{TsangPRL11}. 
Before deriving the QCRB, we briefly review the Fisher information \cite{WisMil10}. 
If we only consider a single unknown parameter $u$ then the Fisher information is a number $F$, and its reciprocal bounds from below the mean-square error (MSE) in any unbiased estimate of the variable. 
For estimation of a set of variables $\{u_j\}$, we have a Fisher information matrix $F_{jk}$, and the bound involves the matrix inverse. 
If now we consider a parameter varying in time $u(t)$ then we replace the square matrix with a function dependent on two arguments, $F(t,t')$. 
This is the case we will be dealing with.

As mentioned previously, we assume the beam has stationary statistics. Consequently $F(t,t')$ can be replaced with $F(t-t')$. 
Moreover, to bound the MSE we can take this Fisher information function to be comprised of a classical and a quantum component,
\beq
\label{Fish-info}
F(t-t')=F^{(C)}(t-t')+F^{(Q)}(t-t') \, .
\eeq
Here $F^{(C)}$ encodes any prior information about the phase, 
while $F^{(Q)}$ is a property of the quantum system(s) from which 
we obtain any further information about the phase. 
The QCRB on the MSE is then given by \cite{TsangPRL11}
\beq
\label{Fish-MSE}
\ex{[\varphi(t)-\breve{\varphi}(t)]^2}\geq F^{-1}(0) \, ,
\eeq
where $\breve{\varphi}(t)$ is any unbiassed estimate of the phase and the inverse of the Fisher information function is defined by
\beq
\label{Fish-inv}
\int_{-\infty}^{\infty} ds \, F^{-1}(t-s)F(s-t') = \delta(t-t') \, .
\eeq
If we then take the Fourier transform of \erf{Fish-inv} then we find $\widetilde{F^{-1}}(\omega)=1/\widetilde{F}(\omega)$. Substituting in \erf{Fish-info} yields
\beq
\label{Fish-FT}
\widetilde{F^{-1}}(\omega)=\frac{1}{\widetilde{F}^{(C)}(\omega)+\widetilde{F}^{(Q)}(\omega)}\, ,
\eeq
and $F^{-1}(0)$ is obtained by integrating over $\omega$:
\beq
\label{Fish-rev}
F^{-1}(0)=\frac{1}{2\pi} \int_{-\infty}^{\infty} d\omega \, \widetilde{F^{-1}}(\omega)\, .
\eeq

For a coherent state beam, the quantum component of the Fisher information is given by $F^{(Q)}(t-t')=4{\cal N}\delta(t-t')$ \cite{DominicPRL13}.
Hence performing a Fourier transform yields $\widetilde{F}^{(Q)}(\omega)=4{\cal N}$.
The classical (prior information) contribution to the Fisher information is given by $F^{(C)}(t-t')=\Sigma^{-1}(t-t')$ \cite{DominicPRL13}, from the assumption that the phase fluctuations are Gaussian. This means that $\widetilde{F}^{(C)}(\omega)=1/S_\varphi(\omega)$ and we can express \erf{Fish-FT} as
\beq
\label{Fish-comp}
\widetilde{F^{-1}} (\omega)=\left[\frac{1}{S_\varphi(\omega)} +4{\cal N}\right]^{-1} \,.
\eeq
Finally, substituting \erf{Fish-comp} into \erf{Fish-rev} gives
\beq \label{QCRB}
{\rm MSE}\ge F^{-1}(0)=\int_{-\infty}^\infty\frac{d\omega}{2\pi}\left[\frac{1}{S_\varphi(\omega)} +4{\cal N}\right]^{-1}\,.
\eeq

\section{Optimal Filtering}
\label{sec-optf}
\begin{figure}[t]
\includegraphics[scale=0.35]{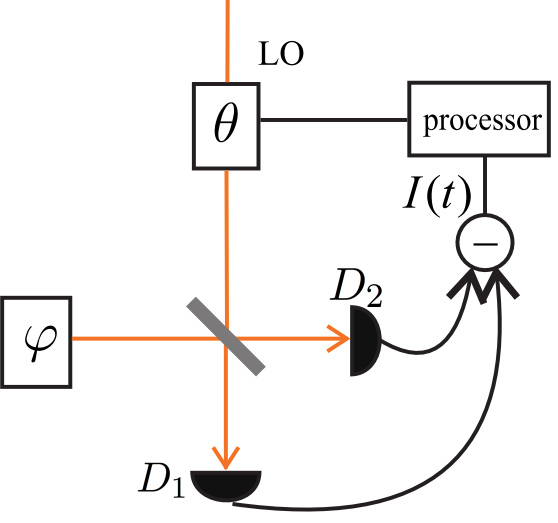}
\caption{The scheme for an adaptive homodyne measurement of the phase of a coherent beam, $\varphi$. $D_1$ and $D_2$ are the photodetectors measuring the two outputs of the 50/50 beam splitter. $I(t)$ is the difference in the photocurrent between $D_1$ and $D_2$. The processor adjusts the phase of the LO, $\theta$, based on $I(t)$.}
\label{adapt-sch}
\end{figure}

Consider an adaptive homodyne measurement scheme to determine the phase, as seen in Fig.~\ref{adapt-sch}.
The quadratures of a beam can be measured by combining it with a strong LO using a 50/50 beam splitter.
The quadrature measurement arises from the difference in the photocurrent $I(t)$ from the outputs of the beam splitter.
The LO also has its own phase shift $\theta$ that may be controlled. 
To measure the phase quadrature, the phase of the local oscillator $\theta$ is chosen to be close to the phase of the beam $\varphi$.
When $\varphi$ is unknown, an adaptive scheme can be used, where $\theta$ is varied during the measurement to track the phase $\varphi$~\cite{Wiseman95}. 

For a coherent beam the expression for the photocurrent is given by \cite{DominicPRA02}
\beq \label{newqI}
I = 2\sqrt{\cal N} \sin \left( {\varphi  - \theta } \right) + \zeta(t)\, ,
\eeq
where $\zeta(t)$ is real classical white noise, satisfying $\langle \zeta(t)\zeta(t')\rangle = \delta(t-t')$.
In the coherent-state case any adaptive scheme which ensures that  $|\theta  - \varphi | \ll 1$ for all time 
(or even for all but a small proportion of time) will be practically as good as one in which $\theta = \varphi$. 
This being the case, we can linearize \erf{newqI} to obtain 
\beq
I \approx 2\sqrt{\cal N} \left( {\varphi  - \theta } \right) + \zeta(t)\, .
\eeq 
It is convenient to add $\theta$ to $I/2\sqrt{\cal N}$ to give the signal
\beq\label{sigex}
r(t) := I/(2\sqrt{\cal N})+\theta \,,
\eeq
where we have scaled the photocurrent to simplify the calculations for both the filtering and the smoothing cases.
Then in the linear approximation the signal is independent of $\theta$, and is
\beq \label{defr}
r(t) \approx \varphi(t)  + n(t) \, .
\eeq
where $n(t) = {\zeta(t)}/{2\sqrt{\cal N}}$. 
The spectrum of the measurement is then given by 
\beq
S_{r}(\omega)=S_{\varphi}(\omega)+S_{n}\,,
\eeq
where the spectrum of the measurement noise is $S_n=1/4{\cal N}$. 
It is important to note that the measurement noise is independent of any stochastic increment in the phase variation.

The optimal estimate of a time-varying phase, or any parameter, is the estimate that minimizes the MSE.
Finding the optimal estimate is typically a difficult problem to solve.
However if we consider the stationary or long-time case, then we can apply Wiener's frequency domain approach to filtering \cite{TSL09, vanTrees1, vanTrees2}. 
The minimum MSE for a signal of the form $r(t)=\varphi(t)+n(t)$, 
where $n(t)$ is Gaussian white noise, is \cite[p.~803]{vanTrees2}
\beq\label{mMSE}
{\rm MSE}_F=S_{n}\int_{-\infty}^\infty \frac{d\omega}{2\pi} \ln \left[1+\frac{S_\varphi(\omega)}{S_{n}}\right]\,,
\eeq
where the subscript $F$ indicates filtering (and we will use $S$ for smoothing). Using $S_n = 1/(4{\cal N})$ and the inequality $\ln(1+x)>x/(1+x)$ for $x>0$, it follows that 
\beq\label{ineq}
{\rm MSE}_F>\int_{-\infty}^\infty\frac{d\omega}{2\pi}\left[\frac{1}{S_\varphi(\omega)} +4{\cal N}\right]^{-1}\,,
\eeq
where the right-hand side is identical to the expression in Eq.~\eqref{QCRB}.
This indicates that the filtered estimate will never attain the QCRB.
Nevertheless, for $x\ll 1$, $\qquad \ln(1+x)\approx x/(1+x)$, so it could be expected that the filtered estimate is close to the QCRB for $4{\cal N}S_\varphi(\omega)$ small.

We consider the case that the phase has a power law spectral density; that is $S_\varphi(\omega)=\kappa^{p-1}/|\omega|^{p}$. 
Substituting in the spectral densities for both the noise and the phase into \erf{mMSE} and using integration by parts, we get

\beq
{\rm MSE}_F=(4{\cal N}\pi)^{-1}p\int_0^\infty\frac{d\omega}{1+\omega^p/\gam}\,,
\eeq
where $\gam=4{\cal N}\kappa^{p-1}$, for $p>1$. 
Solving this integral yields the filtered MSE
\beq \label{filtMSE}
{\rm MSE}_F=[\sin (\pi/p)]^{-1}(4{\cal N}/\kappa)^{-(p-1)/p}\,,
\eeq
for $p>1$.

We can also solve for the QCRB for this specific spectrum using \erf{QCRB}, to obtain 
\beq \label{QFIMSE}
{\rm MSE}  \geq [p \sin (\pi/p)]^{-1}(4{\cal N}/\kappa)^{-(p-1)/p}\, .
\eeq
Note that, unlike the filtered MSE (\ref{filtMSE}), 
no photocurrent linearization assumption is necessary to derive 
\erf{QFIMSE}. Indeed, we do not make any assumptions on 
how the coherent beam is measured.

The filtered estimate has the same scaling with $\cal N$ as the QCRB (\ref{QFIMSE}), but does not attain the QCRB prefactor exactly.
This difference is what was expected from the inequality in Eq.~\eqref{ineq}.
The surprising feature of the result is that the prefactor for filtering diverges from the QCRB linearly in $p$.
However, it is possible to reduce the MSE of the estimate by using the information about the system more effectively, as will be explored in Sec.~\ref{sec-smooth}.

Another interesting feature is that filtering gives a prefactor close to that for the QCRB for $p$ close to $1$ (though both prefactors diverge as $p\to 1$).
That is, filtering gives close to the best estimate, despite using only half of the possible data.
As discussed above, the inequality is due to the inequality $\ln(1+x)>x/(1+x)$, which is close to equality when $x\ll 1$.
Because $x$ corresponds to $S_{\varphi}(\omega)/S_n=4{\cal N}S_{\varphi}(\omega)$, we can expect the results to be close if ${\cal N}S_{\varphi}(\omega)\ll 1$.
This can not be true for all $\omega$, because $S_{\varphi}(0)$ is large regardless of $p$.
However, the MSE depends on an integral over $\omega$.
It turns out that, for $p$ close to $1$, the majority of the contribution to the integral is for values of $\omega$ where ${\cal N}S_{\varphi}(\omega)\ll 1$.
As a result, $\ln(1+x)\approx x/(1+x)$ for most of the contribution to the integral, and the prefactor for filtering is close to the QCRB.
\section{Optimal Smoothing}
\label{sec-smooth}
As noted in Sec.~\ref{sec-optf}, the filtered estimate cannot attain the QCRB for the MSE.
However, smoothing~\cite{TSL09,Wheatley,Yonezawa} can give a better estimate by estimating $\varphi(t)$ using the signal $r(s)$ for $s>t$ as well as $s<t$. 
Since we are now considering twice as much data, the estimate will be more accurate than filtering, and one might expect a factor of $2$ improvement.

We will now show that Wiener's frequency domain approach to smoothing \cite{vanTrees1,vanTrees2,TSL09} achieves the QCRB with coherent states, even allowing an arbitrary spectrum for the phase. 
For a noisy record of the form in \erf{defr}, 
the minimum MSE for smoothing is given by \cite[p.~802]{vanTrees1}
\beq\label{WFunreal}
{\rm MSE}_S=\int_{-\infty}^\infty\frac{d\omega}{2\pi}\left[\frac{1}{S_\varphi(\omega)} +\frac{1}{S_{n}}\right]^{-1}\,.
\eeq
Substituting in for the spectrum of the noise, $S_{n}=1/4{\cal N}$, we arrive at 
\beq 
{\rm MSE}_S=\int_{-\infty}^\infty\frac{d\omega}{2\pi}\left[\frac{1}{S_\varphi(\omega)} +4{\cal N}\right]^{-1}\,,
\eeq
where this is the exact expression in the QCRB given in \erf{QCRB}.

That is to say, the smoothed estimate of the MSE will achieve the lower limit given by the QCRB, for an arbitrary spectrum of the phase. 
This may not necessarily be surprising, as it is the estimate that makes use of all possible information.
However, when we consider the case $S_\varphi(\omega)=\kappa^{p-1}/|\omega|^p$, the filtered error, seen in Sec.~\ref{sec-optf}, diverges from the QCRB by a factor of $p$.
As we have found that smoothing can reach the QCRB, this shows that smoothing provides an unbounded improvement over filtering.
In the $p=2$ case previously considered \cite{Wheatley,TSL09}, smoothing only offered a factor of 2 improvement.

\section{LG Estimation}\label{sec-LG}
Ultimately, we are not only interested in the minimum error in an estimate of the phase, but also 
how to make that estimate. Whilst it is possible to obtain smoothed estimators from the frequency approach without much trouble, it turns out that it is rather difficult to determine the filtered estimator. The problem is that the 
closed-form solution for the filter~\cite[p.~788]{vanTrees1} 
assumes a spectrum $S_\varphi$ that is a rational function. 
It is possible to approximate the spectrum arbitrarily accurately over a given frequency range using rational functions, but more accurate approximations will require more complicated filters.
Moreover, a single approximation cannot be accurate for all frequencies, because $S_\varphi(\omega)=\kappa^{p-1}/|\omega|^p$ for noninteger $p$ will always have different asymptotic scaling than a rational approximation.
It would be necessary to choose the approximation dependent on $\kappa$ and ${\cal N}$ in order to make it accurate over the appropriate range of frequencies.

On the other hand, for even integer $p$, it is possible to easily describe the estimators by formulating the system as an LG estimation problem. Again, we are considering an adaptive homodyne measuring scheme, where the photocurrent is given by \erf{newqI}.
However, we now rescale the linearized photocurrent 
to fit with the convention of LG theory as 
\beq\label{sigex}
y(t) := I+2\sqrt{\cal N}\theta \,.
\eeq
Then the linear approximation is
\beq \label{defy}
y(t)\approx2\sqrt{{\cal N}}\varphi(t) + \zeta(t)\,.
\eeq

To apply LG estimation theory, we consider 
$p=2n+2$, $n\in\mathbb{N}_0$, and define
\begin{align}
\varphi(t) &:= x_n(t) \kappa^{n+1/2}\, , \label{xn} \\
x_{k+1}(t) &:= \int_{-\infty}^t  ds \, x_k( s  )\, , \quad k\in\mathbb{N}_0\, , \label{xrecur} \\
x_0(t) &:= \int_{-\infty}^t dW(s)\, , \label{x0}
\end{align}
where $dW(s)$ is an infinitesimal Wiener increment.
Then it is easy to verify that $\varphi(t)$ is a Gaussian stochastic process with spectrum 
$S_\varphi(\omega) = \kappa^{p-1}/|\omega|^p$ by considering $S_\varphi(\omega)=\int_{-\infty} ^\infty \ex{\varphi(\omega)\varphi(\omega')}d\omega'$.
The system of equations (\ref{defy})--(\ref{x0}) then form what is known as a LG estimation system~\cite{WisMil10}. 

To write the system in standard form~\cite{WisMil10}, 
we define the following vector 
\beq
{\bf x} = (x_0, x_1, \cdots, x_n)^T\, , 
\eeq
and matrices 
\begin{align}
A&=\left( \begin{array}{ccccc}
0 & 0 & 0 & 0 & \\
1 & 0 & 0 & 0 & \\
0 & 1 & 0 & 0 & \cdots \\
0 & 0 & 1 & 0 & \\
 & & \vdots & & \\
\end{array}
\right) , \\
E&=\left( \begin{array}{ccccc}
1 & 0 & 0 & \cdots & 0 \\
\end{array}\right)^T\, , \\
C&=\left( \begin{array}{ccccc}
0 & 0 & 0 & \cdots & \sqrt{\gam} \\
\end{array}\right)\, ,
\end{align}
with $\gam = 4 \mathcal{N}\kappa^{2n+1}$. To be precise: 
$A$ is of dimension $(n+1)\times (n+1)$, and has elements $A_{j,k}=\delta_{j,k+1}$; 
$C$ is of dimension $1\times (n+1)$, and has elements $C_{k}=\sqrt{\gam}\delta_{k,n}$; 
$E$ is of dimension $(n+1)\times 1$, and has elements $E_k=\delta_{k,0}$. 
Then the LG system (\ref{defy})--(\ref{x0}) can be rewritten as 
\begin{align}
d{\bf x}(t) &= A\, \mathbf{x}(t) \, dt +E\, dW(t)\, ,  \label{xdyn} \\
y(t) &=C\, \mathbf{x}(t) + \zeta(t)\, , \label{LGy}
\end{align}
where $dW(t)$ is that appearing in \erf{x0}. 

\subsection{LG Optimal Filtering}
\label{sec-optfLQG}
The optimal estimator that uses information only up to the current time is the solution of the stochastic differential equation \cite{WisMil10}
\beq d\breve{\bf x}(t) = (A-VC^T C)\breve{\bf x}(t) dt +VC^Ty(t) dt\,,\label{dbx}
\eeq
where $V$ is the covariance matrix $\ex{(\breve{\bf x}-{\bf x})^2}$.
This is stochastic because of the white noise in $y(t)$ as per \erf{LGy}. 
This estimator is often called the filtered estimate.

In general, to determine the covariance matrix $V$ one would have to solve a differential matrix Riccati equation. However, we want the stationary, 
or long-time, covariance matrix, which is given by the algebraic matrix Riccati equation~\cite{WisMil10} 
\begin{equation}
\label{eqsol}
0=AV + VA^T+EE^T-VC^TCV\, .
\end{equation}
To confirm the results of Sec.~\ref{sec-optf} we will calculate the MSE by solving the Riccati equation.
Evaluating the $(k,\ell)$ element of the right-hand side of Eq.~\eqref{eqsol} gives
\begin{equation} \label{MRI}
0 = V_{k-1,\ell}+V_{k,\ell-1}+\delta_{k,0}\delta_{\ell,0}-\gam V_{k,n}V_{n,\ell}\, .
\end{equation}
Starting with taking $k=\ell=0$, and noting that $V$ is symmetric, we obtain
$0=1-\gam V_{0,n}^2$,
and so
\begin{equation} \label{V0n}
V_{0,n} = \frac 1{\sqrt{\gam}}\, .
\end{equation}
If we guess a solution of the form
\begin{equation}
V_{k,\ell} = \tilde{V}_{k,\ell} \, \gam^{\alpha(k+\ell)+\beta}\, ,
\end{equation}
then from \erf{MRI} we get, for $k$ and $\ell$ {\it not} both zero, 
\beq
(\tilde{V}_{k-1,\ell}+\tilde{V}_{k,\ell-1})\gam^{\alpha(k+\ell-1)+\beta}  =  (\tilde{V}_{k,n}\tilde{V}_{n,\ell})\gam^{\alpha(k+\ell+2n)+2\beta}\,.
\eeq
For $\tilde{V}_{k,\ell}$ to be independent of $\mu$ we need
\begin{equation}
(2n+1)\alpha +\beta = -1\, .
\end{equation}
In the case where $k$ and $\ell$ \emph{are} both zero, we have already found that \erf{MRI} is satisfied with
$\tilde{V}_{0,n}\, \gam^{n\alpha+\beta}=\gam^{-1/2}$.
Therefore, to obtain $\tilde{V}_{k,\ell}$ independent of $\mu$ we need
\beq\label{ab2}
n\alpha+\beta=-1/2,
\eeq
and $\tilde{V}_{0,n}=1$.
We consequently obtain two equations with two unknowns $\alpha$ and $\beta$,
which have the solution
\beq
\alpha = \beta = -\frac{1}{2n+2}\, .
\eeq
Thus we find that with
\beq  \label{Vklgen}
V_{k,\ell} =
\tilde{V}_{k,\ell} \, \mu^{-(k+\ell+1)/p}\, ,
\eeq
$\tilde{V}_{k,\ell}$ is independent of $\mu$.
Then \erf{MRI} simplifies to a recurrence relation for $\tilde{V}_{k,\ell}$ that is independent of $\mu$
\beq \label{Vtild}
\tilde{V}_{k-1,\ell}+\tilde{V}_{k,\ell-1}+\delta_{k,0}\delta_{\ell,0}-\tilde{V}_{k,n}\tilde{V}_{n,\ell}=0\,.
\eeq
This was solved analytically for $n=0,1,2$, giving solutions
\begin{align}
\tilde{V}&=1 ,\\
\tilde{V}&=\left( \begin{array}{cc}
\sqrt{2} & 1 \\
1 & \sqrt{2} \\
\end{array}
\right) ,\\
\tilde{V}&=\left( \begin{array}{ccc}
2 & 2 & 1 \\
2 & 3 & 2 \\
1 & 2 & 2 \\
\end{array}
\right)\, ,
\end{align}
respectively. For larger $n$ we solved \erf{Vtild} numerically. Some patterns about $\tilde{V}$ are already apparent in the $n=2$ case. 
These matrices, as well as being symmetric about the diagonal, are symmetric about the anti-diagonal. 
Additionally, the top row satisfies $\tilde{V}_{0,k}=\tilde{V}_{0,n-1-k}$ for $k\leq n-1$. 
These patterns persist throughout all calculated matrices for $\tilde{V}$. 

\begin{figure}[t]
\includegraphics[scale=0.30]{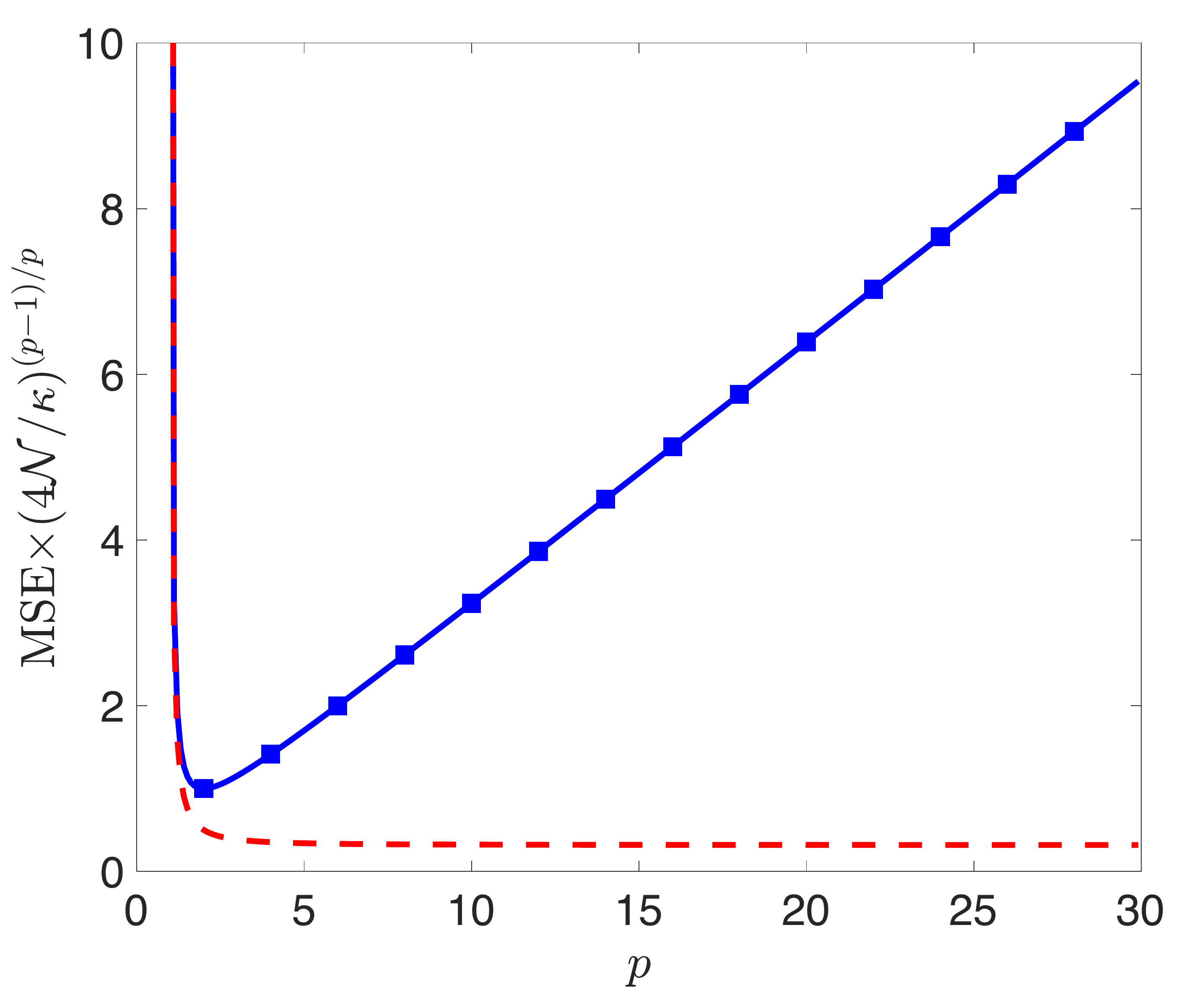}
\caption{(Colour online) The optimal filtered MSE with coherent states using Wiener filtering (blue solid) and the LG model (blue squares) compared to the QCRB (red dashed).}
\label{fig_Vf}
\end{figure}

Substituting $\gam=4 \mathcal{N}\kappa^{2n+1}$ and using \erf{xn} we obtain 
for the achievable MSE 
\beq \label{LGMSE}
{\rm MSE}_F = \kappa^{2n+1} V_{n,n}
= \tilde{V}_{n,n}  ({4\cal N}/\kappa)^{-(p-1)/p}\, ,
\eeq
for $p=2n+2$ even positive integer.
From Fig.~\ref{fig_Vf}, we see that the LG method for calculating the MSE in the phase, whilst limited to even $p$, matches the frequency domain approach of \erf{filtMSE}, as expected.

\subsection{Optimal Smoothing}
\label{sec-smoothLQG}
Smoothing in the LG regime can be considered as a two-filter system \cite{TSL09,Weinert01} whereby the first filter considers all prior data (i.e.~the filtered estimate) and the second filter uses only information after the current time, known as the \emph{retrofiltered} estimate.
In practice, one would need to take all the data first in order to calculate this retrofiltered estimate.
Because the system is completely reversible, we can use exactly the same equations, except reversing the direction of time.
That is,
\beq d\breve{\bf x}_{R}(t) = -(A-V_{R}C^T C)\breve{\bf x}_{R}(t) dt -V_{R}C^Ty(t) dt,\label{dbx2}
\eeq
where $\breve{\bf x}_{R}$ is the retrofiltered estimate. 
Henceforth we will use the subscript $R$ for the retrofiltered quantities, subscript $F$ for the filtered quantities, and subscript $S$ for the smoothed quantities.
This equation needs to be solved backwards in time, which means that numerically the $-dt$ is replaced with a positive increment, and there is no difference from the equations to be solved for the filtered estimate, 
except for the interval over which the data is taken. 
In order to obtain the best possible estimate from the data, we can use both the filtered and the retrofiltered estimate.
This is known as the optimal smoothed estimate,
\beq \label{xs}
\breve{\bf x}_S=V_S(V_F^{-1}\breve{\bf x}_F + V_R^{-1} \breve{\bf x}_R)
\eeq
with smoothed covariance 
\begin{equation}
\label{smooth-err}
V_{S}^{-1}=V_{F}^{-1}+V_{R}^{-1},
\end{equation}
where $V_{F}$ and $V_{R}$ satisfy the stationary filtered and retrofiltered equations
\begin{align}\label{filt}
0&=AV_{F}+V_{F}A^T+EE^T-V_{F}C^TCV_{F}\,,\\
\label{retro}
0&=-AV_{R}-V_{R}A^T+EE^T-V_{R}C^TCV_{R}\,,
\end{align}
respectively \cite{Weinert01}. 
It should be mentioned that we are only considering the stationary solution, eliminating the requirement for initial and final conditions. 

To confirm the results of Sec.~\ref{sec-smooth} we will now determine the smoothed covariance $V_S$. The same approach used for the filtered case was applied to solve for the retrofiltered covariance. It is easy to show that by choosing $[V_{R}]_{k,\ell}=[\tilde{V}_{R}]_{k,\ell}\, \gam^{-(k+\ell+1)/p}$ we arrive at
\begin{equation}
[\tilde{V}_{R}]_{k,\ell}=(-1)^{k+\ell}[\tilde{V}_{F}]_{k,\ell}
\end{equation}
for the solution to the retrofiltered equation.

It can be verified that 
\beq
V_{F}^{-1}=[\tilde{V}_{F}^{-1}]_{k,\ell} \, \gam^{(k+\ell+1)/p}\, ,
\eeq
and similarly for $V_{R}$. Thus
\beq
[V_{S}^{-1}]_{k,\ell}=\left([\tilde{V}_{F}^{-1}]_{k,\ell}+[\tilde{V}_{R}^{-1}]_{k,\ell}\right)\gam^{(k+\ell+1)/p}\, .
\eeq
It can be shown that the smoothed covariance $V_S$ has the form $[V_{S}]_{k,\ell}=[\tilde{V}_{S}]_{k,\ell} \, \gam^{(k+\ell+1)/p}$, where $\tilde{V}_S$ is independent of $\gam$. Then
Eq.~\eqref{smooth-err} simplifies to 
\begin{equation}\label{Vstilde}
\tilde{V}_{S}=\left(\tilde{V}_{F}^{-1}+\tilde{V}_{R}^{-1}\right)^{-1}\,.
\end{equation}
Unlike the filtered estimate, we were able to analytically solve for $[\tilde V_S]_{k,\ell}$ to give 
\beq \label{Vformula}
[\tilde{V}_S]_{k,\ell}=(-1)^{(k-\ell)/2}[p\sin(\pi(k+\ell+1)/p)]^{-1}.
\eeq
Refer to Appendix \ref{app-smooth} for the full derivation.

Like in the filtered case, the smoothed MSE is determined by $[V_{S}]_{n,n}$. Thus 
\beq
{\rm MSE}_S=[p\sin (\pi/p)]^{-1}(4{\cal N}/\kappa)^{-(p-1)/p}, 
\eeq
for even values of $p$. This is exactly the QCRB for the power-law spectrum, \erf{QFIMSE} in Sec.~\ref{sec-optf}, which is what we expect, because the smoothed estimate for LG systems is optimal, and the optimal smoother should attain the QCRB. 



\section{Numerics Without Linearization}
\label{sec-num}
The optimal filtering and smoothing, whilst it does offer insight about the achievable accuracy, is based on a linearization approximation for the photocurrent. 
In this section, we are interested in how the 
linearized theory compares with a simulation of the full nonlinear system. For the cases $p=2$ and $p=4$ we apply the 
optimal filters and smoothers from LG theory to the nonlinear photocurrent. 
We also compare these optimal estimators to a well established method \cite{DominicPRA06,DominicE06} for calculating the phase estimate that does not make a linearization approximation.

The simulation of the phase estimate uses the model of \blk Eqs.~\eqref{xn}--\eqref{x0} with a feedback loop based on the photocurrent, $I(t)$, ensuring that the phase of the LO is equal to the estimate of the time-varying phase, $\theta(t)=\breve\varphi(t)$. 
Where this simulation changes from the model discussed previously is that the photocurrent is not linearized.
That is, rather than \erf{defy} or (\ref{LGy}), we have
\beq\label{yt}
y(t)=I(t)+2\sqrt{\cal N}\theta(t) \, ,
\eeq
where $I(t)$ is given in \erf{newqI}.
The simulation calculates the phase estimate using \erf{dbx} and (\ref{yt}) to then calculate the MSE.

\subsection{Filtered Estimate}
\begin{figure*}[t!]
\includegraphics[scale=0.35]{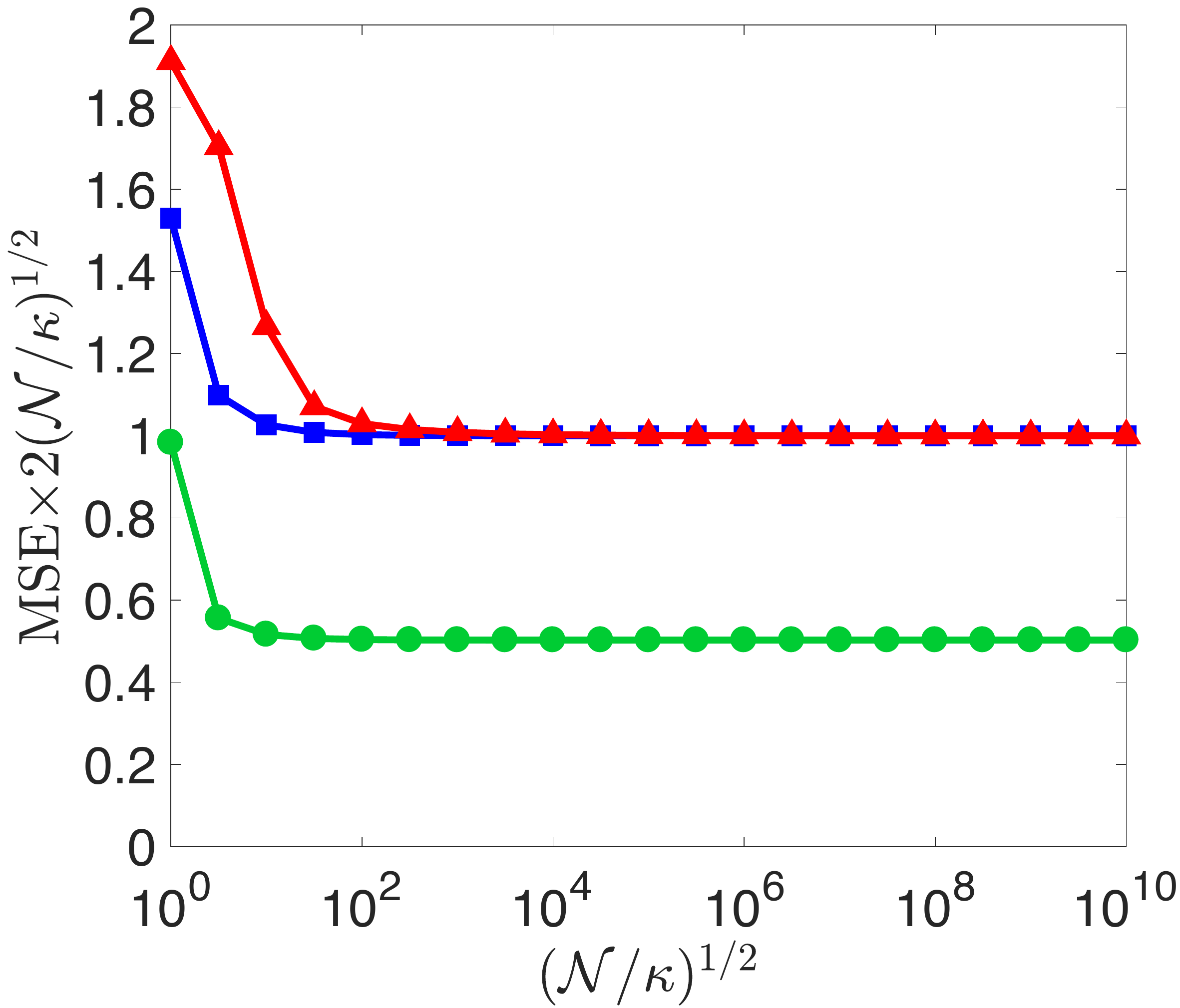}
\includegraphics[scale=0.35]{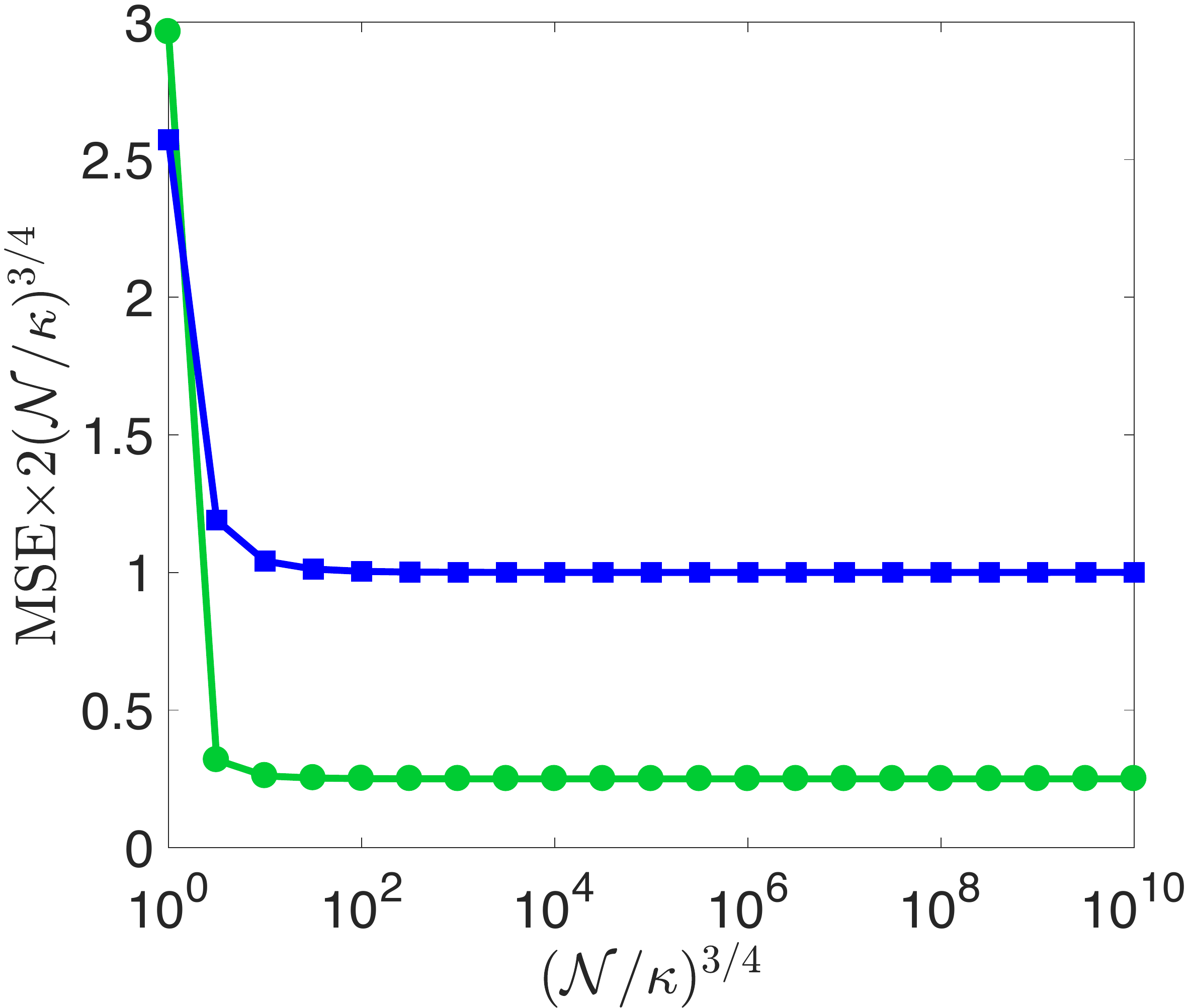}
\caption{(Colour online) The MSE for the simulated filtered estimate (blue squares), the ABC filtered estimates (red triangles) and the optimal smoothed estimates (green circles) with coherent states as a ratio to the predicted filtered values in the asymptotic regime. The left plot is for $p=2$ and the right plot is for $p=4$.}
\label{fig-cohp24}
\end{figure*}

To compare the simulated MSE for the nonlinear system 
to the optimal LG filter for the linear system, 
the ratio (simulated/optimal) was taken. Furthermore, we are interested how this ratio changes with ${\cal N}$. To show this we chose to plot the ratio as a function of the scaling $({\cal N}/\kappa)^{(p-1)/p}$ for ease of comparison with different values of $p$. This is because $({\cal N}/\kappa)^{(p-1)/p}$ is, up to a factor of order unity, the reciprocal of the theoretical asymptotic MSE for any $p$. Thus when this quantity is large (say 100) the filtered phase estimate $\breve\varphi(t)$ can be expected to generally be close to (within $\approx 0.1$ of) the true phase $\varphi(t)$. Because the LO phase $\theta(t)$ is set equal to the filtered phase estimate, this means that, in this regime, the linearization of the photocurrent, needed for the LG theory to be valid, will be a good approximation. 

For the case of both $p=2$ and $p=4$ we see in Fig.~\ref{fig-cohp24} that the simulated MSE does converge to the optimal MSE in the asymptotic limit, as we expected. 
However, as we move closer to ${\cal N}=\kappa$, the ratio gets much worse, increasing by a factor of $1.5$ for $p=2$ and $2.5$ for $p=4$.
This spike corresponds to the linearization of the photocurrent breaking down.
In both cases the asymptotic value was reached, to within an error that is too small to see in the figures, when $({\cal N}/\kappa)^{(p-1)/p} \approx 10^2$.

\subsection{Smoothed Estimate}
We then simulated the smoothed estimate using \erf{xs} and compared it to the optimal LG filter.
We should mention that the expression for the photocurrent (\ref{yt}) still holds for this case, with the LO phase still set as $\theta=\breve\varphi_F$, since the feedback loop must be causal (cannot use any information from the future).
We see for the $p=2$ smoothed estimate, shown in Fig.~\ref{fig-cohp24} (a), the MSE is a factor of $2$ smaller than the filtered estimate in the asymptotic limit, while for $p=4$ in Fig.~\ref{fig-cohp24} (b), 
it is smaller by a factor of $4$. These match predictions, and the lower bound derived from the quantum Fisher information, within about $0.6\%$. 
Moving closer to ${\cal N}/\kappa=1$ we still observe the spike due to the linearization breaking down. 
In the case of $p=2$ the size of the spike is approximately a factor of $2$ larger.
However, looking at $p=4$, it has increased by a factor of $12$, resulting in the smoothed estimate actually performing worse than the filtered. 
Thus, whilst in the asymptotic limit smoothing can offer a large improvement over filtering, there is a diminishing improvement when the linearization breaks down, to the point where the filtered estimate will outperform the smoothed estimate.

\subsection{ABC Method}
Since the linearization of the photocurrent is clearly not valid in the non-asymptotic regime, an obvious solution is to use a method that does not rely on a linearization.
From such a method we might expect better results when $({\cal N}/\kappa)^{(p-1)/p}={\cal O}(1)$.
The ``ABC" method, as we will call it, introduced by Berry and Wiseman \cite{DominicPRA02,DominicPRA06,DominicE06}, uses two functions of the photocurrent record, $a$ and $b$, to calculate the time-varying phase given by, 
\begin{align}
a(t)&=\int_{-\infty}^{t} du\, e^{\chi(u-t)}e^{i\theta} I(u)\, ,\\
b(t)&=-\int_{-\infty}^{t} du\, e^{\chi(u-t)} e^{2i\theta} \, ,
\end{align}
with $1/\chi$ a time constant. The estimate of the phase at time $t$, which is also used for $\theta(t)$, is given by 
\beq
\breve{\varphi} (t)=\text{arg}[c(t)]\, , \qquad c(t)=a(t)+\chi b(t)a^*(t)\, .
\eeq
Note that we have used lower case $a$, $b$ and $c$ as opposed to the capitals used in Ref.~\cite{DominicPRA02,DominicPRA06,DominicE06} so as not to confuse the reader with the previous matrices $A$ and $C$. We performed a simulation using this model for the phase.

In the case of $p=2$, the optimal $\chi$ is known to be $\chi=\sqrt{\gam}$ \cite{DominicPRA02}. In this case, convergence of the ABC MSE to the optimal LG filter MSE in the asymptotic limit can be seen in Fig.~\ref{fig-cohp24}. However, the ABC method  reached its asymptotic value, to within error too small to see in the figures, for $({\cal N}/\kappa)^{(p-1)/p} \approx 10^3$, 
which is slower convergence than the LG filter.
Unexpectedly when the linearization breaks down, i.e.\ ${\cal N}/\kappa = {\cal O}(1)$, the ABC estimate performs worse compared to the optimal estimate, by about 30\%. 
Furthermore, the ABC method performed even worse when $p=4$. 
For long times the variance in the estimate tended to become larger without limit and could not be shown in Fig.~\ref{fig-cohp24}.
We were able to obtain bounded results by introducing a low-frequency cutoff to the correlations, similar (though 
not identical) to that of Ref.~\cite{DinaniARX}. However, when this was done, the results were sensitive to the cutoff. This can be understood analytically, as shown in Appendix~\ref{app}.



\section{Conclusion}
In this paper we investigated the estimation of a time-varying phase of a coherent beam using an adaptive homodyne scheme. 
We consider a phase with time-invariant Gaussian statistics with a power-law spectral density, with exponent $-p$. One can derive the Quantum Cram\'{e}r-Rao Bound, an analytical, asymptotically (in intensity) achievable, bound for the mean-square error. 
This result for coherent beams is an important benchmark against which to judge any quantum advantage.
In the regime where we can linearize the photocurrent, the filtered MSE was found to achieve the same scaling as the QCRB, but the prefactor could not be achieved.
However, when we applied smoothing we found that it achieved the QCRB for arbitrary phase spectrums. 
When considering a power-law spectral density, we observed an improvement greater than a factor of $2$ for $p>2$ over the filtered error.
In fact this improvement increased without bound by a factor of $p$.

To investigate the system in the regime where the linearization is not a good approximation, 
we performed numerical simulation  for both $p=2$ and $p=4$, for both filtering and smoothing. 
In order to perform these simulations, we remodelled the system as an LG estimation problem.
In all cases we observed convergence to the LG theory in the asymptotic limit.
However when ${\cal N}/\kappa = {\cal O}(1)$ the linearization of the photocurrent broke down and the simulation and LG optimal MSE diverged.
We then tested the alternate ``ABC'' method \cite{DominicPRA02,DominicPRA06,DominicE06} that was not based upon a linearized theory. 
For the case of $p=2$ we found that for small $({\cal N}/\kappa)^{1/2}$ the results were worse than the optimal LG filter, but still converged to the same MSE in the asymptotic limit.
Furthermore, for the case of $p=4$, the ABC method could not provide a converging solution.

Observing the unbounded improvement smoothing can offer over filtering experimentally would be an interesting topic for further study.
Given these results, it is also natural to ask if the LG estimators can give improved results, compared to the ABC method, for the squeezed states as well.
Surprisingly, when we tested this estimator for squeezed states, we found it gave very poor results.
Even a small amount of squeezing dramatically increased the MSE above that for the coherent state, instead of decreasing it. 
Thus it is an open problem to find an estimator that does provide improved performance for squeezed states.

\acknowledgments
HMW is supported by the ARC Centre of Excellence Grant No.\ CE110001027. DWB is funded by a Discovery Project (DP160102426). HTD is funded by the Fondecyt-Postdoctrodo Grant No. 3170922.

\appendix
\section{Derivation of the smoothed variance}\label{app-smooth}
To solve a Riccati equation of the form \erf{Vtild}, one can construct a block matrix 
\beq
Z=\begin{pmatrix}
A^T & -\tilde C^T \tilde C \\
-EE^T & -A
\end{pmatrix} \,,
\eeq
with dimensions of $2n+2=p$.
Using the earlier definitions of $A$, $\tilde C$ and $E$ then we can write the elements of $Z$ as 
\beq
Z_{jk} = \begin{cases}
\delta_{j+1,k} & j\le n \\
-\delta_{p,k} & j=n+1 \\
-\delta_{1,k} & j=n+2 \\
-\delta_{j-1,k} & j>n+2
\end{cases}
\eeq
where we are now numbering the rows/columns from $1$ as opposed to $0$.
If we construct a matrix of size $2n+2$ by $n+1$ from the eigenvectors of $Z$, represented in block-matrix notation as
\beq
\begin{pmatrix} Y \\ X \end{pmatrix}\, ,
\eeq
then the solution to the Riccati equation is given by 
\beq
\tilde V_F=XY^{-1} \,.
\eeq
If we solve for the eigenvalues we get
\beq
\lambda_k = i e^{\pi i (2k-1)/p}\,,
\eeq
and can define the matrices
\beq
Y_{jk} = \lambda_k^{j-1}, \quad X_{jk} = \frac 1{(-\lambda_k)^j}\, .
\eeq
We do not have an explicit analytic solution for $X^{-1}$, so we determined it numerically to determine the filtered covariance $V_F$.

We can similarly solve the retrofiltered case using a matrix 
\beq
Z'=\begin{pmatrix}
-A^T & -\tilde C^T \tilde C \\
-EE^T & A
\end{pmatrix} \,,
\eeq
with the elements of $Z'$ given by
\beq
Z'_{jk} = \begin{cases}
-\delta_{j+1,k} & j\le n \\
-\delta_{p,k} & j=n+1 \\
-\delta_{1,k} & j=n+2 \\
\delta_{j-1,k} & j>n+2 \, .
\end{cases}
\eeq
Again we construct a matrix with columns corresponding to the eigenvectors of $Z'$
\beq
\begin{pmatrix} Y' \\ X' \end{pmatrix}\, ,
\eeq
and the solution for the retrofiltered covariance matrix is 
\beq
\tilde V_R=X'(Y')^{-1}\,.
\eeq
Finding the eigenvectors of $Z'$ gives $Y'$ and $X'$ as 
\beq \label{firstform}
Y'_{jk} = (-\lambda_k)^{j-1}, \quad X'_{jk} = -\frac 1{\lambda_k^j}\,.
\eeq

These eigenvectors look very similar to the eigenvectors for the filtered case. To see the similarity, we create another set of eigenvectors by multiplying each column by $-\lambda_k$ to give
\beq
Y'_{jk} = (-\lambda_k)^{j}, \quad X'_{jk} = \frac 1{\lambda_k^{j-1}}\,.
\eeq
Since the covariance matrix is real, we can obtain the same covariance matrix using the complex conjugates of $Y'$ and $X'$.
Using $\lambda_k^*=1/\lambda_k$ the new matrices are
\beq\label{modified}
Y'_{jk} =\frac 1{ (-\lambda_k)^{j}}, \quad X'_{jk} = {\lambda_k^{j-1}}\,.
\eeq
We can see that $X'=Y$ and $Y'=X$, or $\tilde V_R=X^{-1}Y$ which is the inverse of $\tilde V_F$, implying $(\tilde V_F)^{-1}=\tilde V_R$. 

It is straightforward to show from Eqs.~\eqref{filt} and \eqref{retro} that the inverse of $\tilde V_F$ satisfies the same equation as $\tilde V_R$, except flipped on the anti-diagonal.
This is enough to show that $\tilde V_F^{-1}$ is the pertranspose of $\tilde V_R$, as well as being equal to $\tilde V_R$.
As a result, $\tilde V_F$ and $\tilde V_R$ are bisymmetric.

To determine the smoothed variance we wish to determine 
\beq
\tilde V_S= (XY^{-1}+X'(Y')^{-1})^{-1}\,,
\eeq
which can be rewritten as
\beq\label{VSXY}
\tilde V_S=Y(X+X'(Y')^{-1}Y)^{-1}\,.
\eeq
If we then construct a matrix
\beq \label{block}
\begin{pmatrix} X & -X'\\
Y & Y'
\end{pmatrix}
\eeq
and take its inverse, the upper-left corner of the resulting matrix is the term from \erf{VSXY}, $(X+X'(Y')^{-1}Y)^{-1}$.
This means that if we can take the inverse of this block matrix, we can determine $\tilde V_S$ without explicitly inverting $Y'$.
Using the formula for $\lambda_k$, we have $\lambda_{k+n+1}=-\lambda_k$. Therefore, the rows of the block matrix \eqref{block} are given by the same equation on both the left and the right side. We can then turn this matrix into a form similar to a Fourier transform matrix. Thus we define a new matrix,
\beq
T=\begin{pmatrix} W & W' \\
V & V'
\end{pmatrix}
=
\begin{pmatrix} J & 0 \\
0 & I
\end{pmatrix}
\begin{pmatrix} X & -X' \\
Y & Y'
\end{pmatrix}\,,
\eeq
where $I$ is an identity matrix, and $J$ is an anti-diagonal matrix with entries
\beq
J_{jk} = (-1)^k \delta_{j,n+2-k}\,.
\eeq
We use $X'$ and $Y'$ as given by the original form in Eq.~\eqref{firstform}, not the modified form in Eq.~\eqref{modified}.
The resulting $W$ matrix has elements 
\begin{align}
W_{jk} &= \sum_\ell J_{j\ell} X_{\ell k} \nn
&= \sum_{\ell} (-1)^\ell \delta_{j,n+2-\ell} (-\lambda_k)^{-\ell} \nn
&= \lambda_k^{j-(n+2)}.
\end{align}
Similarly $W'$ is given by
\begin{align}
W'_{jk} &= -\sum_\ell J_{j\ell} X'_{\ell k} \nn
&= \sum_{\ell} (-1)^\ell \delta_{j,n+2-\ell} \lambda_k^{-\ell}\nn
& = (-\lambda_k)^{j-(n+2)}\nn
&= \lambda_{k+n+1}^{j-(n+2)} \, .
\end{align}
As a result, the formulas for each block of $T$ are consistent, and
we can describe the entire matrix $T$ by the same formula,
\beq
T_{jk}=\lambda_k^{j-(n+2)}\,.
\eeq
The inverse of $T$ is well known and has the matrix elements
\beq
(T^{-1})_{jk} = \frac 1p \lambda_j^{(n+2)-k}\,.
\eeq

Now we have
\beq
\begin{pmatrix} X & -X' \\
Y & Y'
\end{pmatrix}^{-1} = T^{-1} \begin{pmatrix} J & 0 \\
0 & I
\end{pmatrix}\,.
\eeq
Therefore, to get the upper-left block of the inverse, we need multiply the upper-left block of $T^{-1}$ by $J$.
That gives the simple answer
\begin{align}
[(X+X'Y'^{-1}Y)^{-1}]_{jk} &= \frac 1p \sum_\ell \lambda_j^{(n+2)-\ell}(-1)^k \delta_{\ell,n+2-k}\nonumber \\
&= \frac 1p (-\lambda_j)^k\,.
\end{align}
To get the final answer, we need to just multiply by $Y$
\begin{align}
[\tilde V_S]_{jk}&=[Y(X+X'Y'^{-1}Y)^{-1}]_{jk} \nonumber \\
&= \sum_{\ell=1}^{n+1} \lambda_{\ell}^{j-1} \frac 1p (-\lambda_\ell)^k \nonumber \\
&= \frac 1p (-1)^k \sum_{\ell=1}^{n+1} \lambda_\ell^{j+k-1}\nonumber \\
&= \frac 1p (-1)^k i^{j+k-1} \sum_{\ell=1}^{n+1} e^{\pi i (2\ell-1)(j+k-1)/p}\nonumber \\
&= \frac 1p (-1)^k i^{j+k-1} e^{\pi i (j+k-1)/p} \frac{e^{\pi i (j+k-1)}-1}{e^{2\pi i (j+k-1)/p}-1}\nonumber \\
&= \frac 1p (-1)^k i^{j+k-1} \frac{e^{\pi i (j+k-1)}-1}{e^{\pi i (j+k-1)/p}-e^{-\pi i (j+k-1)/p}}\nonumber \\
&= \frac{1-e^{\pi i (j+k-1)}}2 \frac { i^{j-k}}{p\sin(\pi (j+k-1)/p)}\,.
\end{align}

Now if we have $j+k-1$ even, or equivalently $j-k$ odd, then $e^{\pi i (j+k-1)}=1$ so we get zero.
If $j-k$ is even, then we have the result
\beq
\frac { (-1)^{(j-k)/2}}{p\sin(\pi (j+k-1)/p)}.
\eeq
In this expression we are taking $j$ and $k$ numbered from $1$, in contrast to the numbering from $0$ in the body of the paper.
Switching to numbering from $0$ gives the expression in Eq.~\eqref{Vformula}.
Taking $j=k=n+1$, we get
\beq
\frac 1{p\sin(\pi (p-1)/p)} = \frac 1{p\sin(\pi/p)}
\eeq
as required.\\

\section{Derivation of MSE for simple estimator}
\label{app}
As stated in the main text, the ABC estimator works 
for $p=2$, but for $p=4$ the simulations give divergent results.
Here we consider phase estimation with a coherent beam, where the high-frequency phase spectrum has an inverse power $p$, with $p$ a positive even integer. We wish to consider 
estimation using the method in Sec.~IV of Ref.~\cite{DinaniARX}. This divergence can be fixed by introducing decay in the phase dynamics. Specifically, we modify Eqs.~(\ref{xrecur}) and (\ref{x0}) by 
\begin{align}
x_{k+1}(t) &= \int_{-\infty}^t du\,e^{\lambda_{k+1}(u-t)} x_k( u  ), \quad k\in\mathbb{N}, \\
x_0(t) &= \int_{-\infty}^t dW(u)\,e^{\lambda_0(u-t)} .
\end{align}
As stated, this gives convergent results for the phase uncertainty, but the results are 
sensitive to the exact values of $\lambda_k$ used.

The behaviour just described can be predicted in a simplified linearized theory. 
Using the theory in Ref.~\cite{DominicPRA02}, it was shown that for a coherent state, the phase estimate $\breve{\varphi}(t)$ using the ABC method can be approximated by
\beq
\breve{\varphi}(t) = \chi \int_{-\infty}^{t} du\,\left[\theta(u)+\frac{I(u)}{2\sqrt{\cal N}} \right] e^{\chi(u-t)} \, .
\eeq
Linearizing the photocurrent reduces this equation to
\begin{equation}
\breve\varphi = \frac{\chi}{2\sqrt{\cal N}}\int_{-\infty}^t du\,e^{\chi(u-t)} y(u)  \, ,
\end{equation}
and from this it is straightforward to show that, 
for $p=4$, the predicted MSE diverges if $\lambda_1=\lambda_0 = 0$, 
while if we take $\lambda_0 = \lambda\neq 0,$ $\lambda_1=0$ then 
convergent results can be obtained, albeit dependent heavily on $\lambda$.

In both cases, the MSE is given by
\begin{widetext}
\begin{align}
\langle (\breve\varphi - \varphi)^2 \rangle &= \left\langle \left( 
\chi \int_{-\infty}^t du\,e^{\chi(u-t)} \varphi(u) + \int_{-\infty}^t dW(u) \,e^{\chi(u-t)} - \varphi(t)
\right)^2 \right\rangle \nn
&=  \left\langle \left( 
\chi \int_{-\infty}^t du\, e^{\chi(u-t)} \varphi(u) \, - \varphi(t)
\right)^2 \right\rangle +  \left\langle \left( 
\int_{-\infty}^t dW(u)\, e^{\chi(u-t)} \right)^2 \right\rangle \nn
&=  \left\langle \left( 
\chi \int_{-\infty}^t du\, e^{\chi(u-t)} [\varphi(u)-\varphi(t)] \, 
\right)^2 \right\rangle + \frac 1{2\chi}\nn
&= 
\chi \int_{-\infty}^t du_1 \int_{-\infty}^t du_2 \, e^{\chi(u_1+u_2 -2t)} \left\langle[\varphi(u_1)-\varphi(t)][\varphi(u_2)-\varphi(t)]
 \right\rangle + \frac 1{2\chi}\label{evih} .
\end{align}

First consider the divergent case, with no cutoff. Then the expectation value 
in \erf{evih} evaluates to
\begin{align}
\left\langle[\varphi(u_1)-\varphi(t)][\varphi(u_2)-\varphi(t)] \right\rangle &=
\kappa^3\int_{u_1}^t dv_1 \int_{u_2}^t dv_2 \, \left\langle x_0(v_1) x_0(v_2) \right\rangle \nn
&= \kappa^3 \int_{u_1}^t dv_1 \int_{u_2}^t dv_2 \, \left\langle \int_{-\infty}^{v_1} \int_{-\infty}^{v_2} dW(w_1) dW(w_2) \right\rangle \nn
&=\infty.
\end{align}
Hence, if there is no damping in the phase variation, the MSE diverges for this estimator.
If we instead introduce a frequency cutoff by setting
$dx_0 = -\lambda x_0 + dW$, then we get
\begin{align}
\left\langle[\varphi(u_1)-\varphi(t)][\varphi(u_2)-\varphi(t)] \right\rangle &= \kappa^3 \int_{u_1}^t dv_1 \int_{u_2}^t dv_2 \, \left\langle \int_{-\infty}^{v_1} \int_{-\infty}^{v_2} dW(w_1) dW(w_2) \, e^{\lambda(w_1+w_2-v_1-v_2)}  \right\rangle \nn
&= \kappa^3 \int_{u_1}^t dv_1 \int_{u_2}^t dv_2 \, \int_{-\infty}^{\min(v_1,v_2)} dw \, e^{\lambda(2w-v_1-v_2)}  \nn
&= \kappa^3 \int_{u_1}^t dv_1 \int_{u_2}^t dv_2 \, \frac 1{2\lambda} e^{-\lambda|v_1-v_2|} \nn
&= \frac{\kappa^3}{2\lambda^3} \left( e^{\lambda (u_1-t)} + e^{\lambda (u_2-t)} - e^{\lambda(-|u_1-u_2|)} - 1 + 2\lambda t - 2\max(u_1,u_2) \lambda \right).
\end{align}
\end{widetext}
Using this expression, a simple integral gives the MSE for the estimator as
\begin{equation}
\langle (\breve\varphi - \varphi)^2 \rangle = \frac {\kappa^3}{2\lambda\chi^3(\lambda+\chi)} + \frac 1{2\chi},
\end{equation}
which shows a sensitive dependence on $\lambda$ as was found numerically.

\end{document}